\documentclass[%
 reprint,
 amsmath,amssymb,
 aps,
]{revtex4-1}

\usepackage{graphicx}
\usepackage{dcolumn}
\usepackage{bm}
\usepackage{xcolor}

\begin{document}

\preprint{APS/123-QED}

\title{Direct Observation of Ion Micromotion in a Linear Paul Trap}

\author{Liudmila A. Zhukas}
\email[]{Corresponding author: lzhukas@uw.edu}
\affiliation{Department of Physics, University of Washington, Seattle 98195, USA}

\author{Maverick J. Millican}
\affiliation{Department of Physics, University of Washington, Seattle 98195, USA}

\author{Peter Svihra}
\affiliation{Department of Physics, Faculty of Nuclear Sciences and Physical Engineering, Czech Technical University, Prague 115 19, Czech Republic}
\affiliation{Department of Physics and Astronomy, 
School of Natural Sciences, 
University of Manchester, Manchester M13 9PL, United Kingdom}

\author{Andrei Nomerotski}
\affiliation{Physics Department, Brookhaven National Laboratory, Upton, NY 11973, USA}

\author{Boris B. Blinov}
\affiliation{Department of Physics, University of Washington, Seattle 98195, USA
}

\date{\today}

\begin{abstract}
In this paper, direct observation of micromotion for multiple ions in a laser-cooled trapped ion crystal is discussed along with a novel measurement technique for micromotion amplitude. Micromotion is directly observed using a time-resolving, single-photon sensitive camera that provides both fluorescence and position data for each ion on the nanosecond time scale. Micromotion amplitude and phase for each ion in the crystal are measured, allowing this method to be sensitive to tilts and shifts of the ion chain from the null of the radiofrequency quadrupole potential in the linear trap. Spatial resolution makes this micromotion detection technique suitable for complex ion configurations, including two-dimensional geometries. It does not require any additional equipment or laser beams, and the modulation of the cooling lasers or trap voltages is not necessary for detection, as it is in other methods. \end{abstract}

\maketitle

\section{Introduction}
The motion of trapped ions in radiofrequency (RF) traps can be described as a superposition of a slower harmonic motion, called the secular motion, and a faster, driven motion at the frequency of the applied electric field, called micromotion. Excess micromotion may lead to significant frequency modulation of the laser light interacting with the ions, reducing the efficiency of laser cooling~\cite{Cirac} and quantum logic gates~\cite{Gaebler16}. Second-order Doppler shifts caused by micromotion can adversely affect trapped ion clocks~\cite{Keller2015}. Thus, detection, monitoring, and compensation of excess micromotion is crucial for high-precision applications such as quantum computing and frequency standards. Micromotion detection and compensation is especially important in chip-scale surface-electrode ion traps because the ions are close to the trap surfaces and, thus, are more susceptible to the stray DC electric fields that shift the trapping position away from the RF null and cause excess micromotion. 

Several techniques for micromotion detection have been developed, including the photon correlation technique and micromotion sideband spectroscopy~\cite{Berkeland98,Tan2019}, high-finesse optical cavity~\cite{Chuah2013}, and parametric resonance detection~\cite{Ibaraki2011}. All methods for micromotion detection listed above are indirect, require additional optical setup, and typically rely on ion fluorescence rate or spectrum changes caused by the ion motion. Another technique~\cite{Keller2015} provides stroboscopic micromotion measurements of multiple ions with an intensified CCD camera using a specially designed pulse generator for imagining synchronized with the trap RF source. This method offers approximately 10~ns time resolution, limited by the trigger pulse duration, and thus, it may not provide enough positional data in the case of lower trap frequencies.

Here the direct observation of the trapped ion micromotion is presented using a time-resolving, single-photon sensitive camera that provides both fluorescence and position data for each ion on a nanosecond time scale. With this technique, both the micromotion amplitude and phase for each ion in the crystal can be measured individually. This can provide information about the tilts and shifts of the ion chain from the null of the radiofrequency quadrupole potential in a linear trap. The high spatial resolution makes this method suitable for micromotion studies in more complex ion configurations, including two-dimensional geometries~\cite{Ivory2020}. Another advantage of this method is its simplicity: no additional equipment, such as time-to-digital converters (TDCs) or high-finesse cavities are required, and no modulation of the trap voltages or lasers is necessary. 

The remainder of the text is organized as follows: in Section \ref{sec:setup} the experimental setup and methods are described in detail. This includes the linear RF trap and the Timepix3 fast camera used for measurement. Section \ref{sec:analysis} provides information on the offline post-processing of the data. Then in Section \ref{sec:results} the results of measured ion micromotion and proper compensation are presented, followed by the team's conclusions in Section \ref{sec:conclusions}.

\section{Setup and Methods}
\label{sec:setup}

\begin{figure}[ht]
    \includegraphics[width=8.5cm]{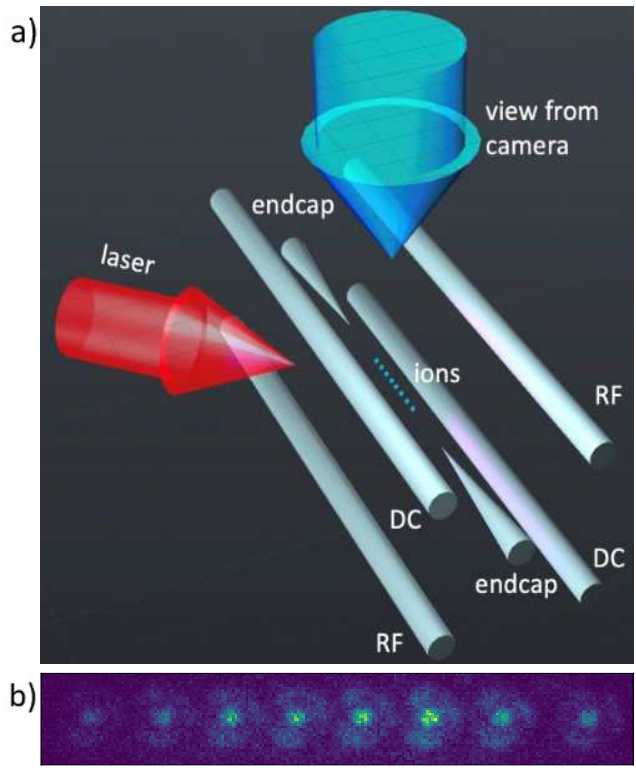}
    \centering
    \caption{\label{fig:Fig1}
    (a) Diagram of the experimental setup with ions in a linear RF trap formed by four rods and two needle endcap electrodes. DC bias voltages may be applied to any of the four rods to compensate the micromotion. The cooling laser (red arrow) is applied at 45 degrees to the trap axis and the ion string is imaged from above. (b) Image of eight Ba$^+$ ions taken with the Tpx3Cam camera over a 20~s integration period. The ion brightness variation is due to the Doppler cooling beam profile. The distance between the middle ions is approximately 5~$\mu$m.
}
    \label{fig:setup}
\end{figure}

An 8-ion chain of Ba$^{+}$ ions is stored and Doppler-cooled in a ``five-rod" linear RF trap~\cite{Dietrich2010}, schematically shown in Figure \ref{fig:setup}a. The ions are laser cooled on the $6S_{1/2}-6P_{1/2}$ transition near 493~nm, a process which produces the fluorescent photons used in the measurement of micromotion. The cooling laser is directed at 45 degrees to the axis of the trap. A 650~nm laser is used to repump ions from the long-lived $5D_{3/2}$ metastable state. The ion fluorescence is collected by a lens with numerical aperture of 0.20 along the direction perpendicular to both the trap axis and the cooling laser. The trap RF frequency is $\Omega=\ 2\pi\cdot18.2516$~MHz at approximately 400 Vpp applied to the pair of the trap rods labelled ``RF" in the Figure 1a. A ``squeeze" voltage of 6 V is applied to the other pair of rods (labelled ``DC" in Figure 1a) to lift the secular frequency degeneracy. The endcap electrodes are at 130 V, resulting in trap secular frequencies ($\omega_x, \omega_y, \omega_z) = 2\pi(1.27,0.98,0.25)$~MHz. Adjustable DC bias voltages may be applied to any of the trap rods to intentionally shift the ions away from the RF null and induce micromotion, or to compensate excess micromotion as the ions are shifted toward the RF null.

The ions are imaged with a single-photon sensitive camera, Tpx3Cam~\cite{timepixcam, tpx3cam, ASI}, with high spatial and temporal resolution.  Figure \ref{fig:setup}b shows the image of eight Ba$^+$ ions in the trap as a two-dimensional distribution of their positions measured by the camera. The camera technology allows direct position detection of single photons on a time scale that is much faster than the 54.79~ns period of the ion micromotion. Multiple photon source position measurements taken over the period of micromotion provide displacement data for individual ions to be analyzed and mapped. Fitting position data for each ion over time to a sinusoidal function gives a direct estimate of the ion micromotion amplitude and phase.

The Tpx3Cam camera has an optical sensor with high quantum efficiency (QE) \cite{Nomerotski2017} which is bump-bonded to Timepix3 \cite{timepix3}, a readout chip with 256$\times$256 pixels of 55$\times$55 $\mu$m$^2$. The electronics in each pixel process incoming signals to measure their time of arrival (ToA) if a predefined threshold with 1.56~ns precision is crossed. The information about time-over-threshold (ToT) is stored together with ToA as time codes in a memory inside the pixel. The ToT information is related to the deposited energy in each pixel.
The Timepix3 readout is data driven so the signal storage and readout proceed only after the pixel signal exceeds the threshold. The pixel dead time is 475~ns~+~ToT allowing for multi-hit functionality in each pixel, independent from the other pixels. The SPIDR readout system supports fast, 80 Mpix/sec bandwidth \cite{spidr}.

The camera was calibrated to equalize the response of all pixels by adjusting the individual pixel thresholds. After this procedure, the effective threshold to fast light flashes is $600-800$ photons per pixel, depending on the wavelength. A small ($\approx0.1\%$) number of hot pixels were masked, i.e. pre-programmed in Timepix3 to stay inactive during the data collection. This prevents saturating the readout bandwidth with useless data since the pixel deadtime is only a few microseconds.

The camera  can  also  accept  and  time  stamp  an  external  pulse,  independently  of  the Timepix3 operation.  In our experiments, these pulses were produced by a generator with frequency of about 100 kHz providing an additional time reference for registered photons. The granularity of the time-stamping for the external pulses is 0.26 ns.

For the single photon  operation, the signal is amplified with the addition of an image intensifier. The image intensifier is a vacuum device comprised of a photocathode followed by a micro-channel plate (MCP) and fast scintillator P47. The hi-QE-green photocathode in the intensifier (Photonis \cite{Photonis}) has QE of about 20$\%$ at 493~nm and dark count rate of about 200 Hz over the full area (18 mm diameter). The MCP in the intensifier has an improved detection efficiency close to 100$\%$. The intensifier is packaged with an integrated power supply and relay optics to project the light flashes from the intensifier output window directly on to the optical sensor in the camera.
Similar configurations of the intensified Tpx3Cam were used previously for trapped ion imaging \cite{Zhukas2020}, characterization of quantum networks \cite{Ianzano2020, Nomerotski2020}, quantum target detection \cite{Yingwen2020, Svihra2020}, single photon counting \cite{Keach2020} and lifetime imaging \cite{Sen2020} studies.

\section{Data Analysis}
\label{sec:analysis}

After sorting data by time, the pixels are grouped into ``clusters" using a recursive algorithm. Clusters are small collections of pixels adjacent to each other that were triggered within a predefined time window of 300~ns. Each cluster corresponds to an individual photon.
Since all hit pixels measure ToA and ToT independently and provide the position information, it can be used for centroiding to determine the coordinates of single photons. The ToT information is used to form a weighted average, giving an estimate of the x, y coordinates for the incoming single photon. The timing of the photon is estimated by using the recorded ToA of the pixel with the largest ToT in the cluster. The above ToA is then adjusted for the so-called time-walk, an effect caused by the variable pixel electronics time response, which depends on the amplitude of the input signal \cite{Turecek_2016, tpx3cam}. With this correction a 2~ns time resolution (rms) has been previously achieved for single photons \cite{Ianzano2020}.

Data was collected for 300 seconds. The average photon detection rate per ion is approximately 1000/s; however, due to the nonuniform cooling laser beam profile, the emission rate varies between the ions.  

\begin{figure}[ht]
    \centering
    \includegraphics[width=8.5cm]{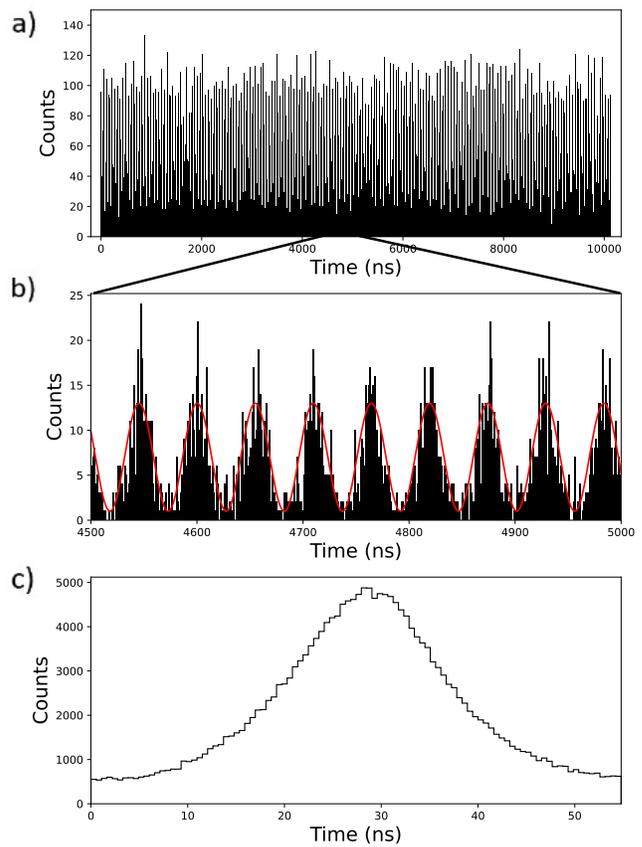}
    \centering
    \caption{(a) Distribution of the time difference between the fluorescent photons and the periodic external signal for the whole period of 10,136 ns. 
    (b) Zoomed-in section of the same distribution for a subset of nine oscillations. The oscillations are well described with a sine function with a period of 54.79~ns (red line), which is exactly the period of the trap RF.
    (c) Time distribution of the entire data set in (a) after applying the data folding procedure over the oscillation period as described in the text. The resulting distribution represents the probability of a photon emission during the micromotion period.}
    \label{fig:deltat}
\end{figure}

The photon emission rate for ions in a trap oscillates with the RF period because of the varying Doppler shift as the velocity of the ions change with respect to the laser beam due to micromotion. During data collection, external reference signals were provided to the camera by a pulse generator that was running synchronously with the trap RF, generating a pulse every 185th RF period, at a rate of approximately 100~kHz.
This time reference was used to measure the oscillations by forming the time difference between the photon emission time and reference signal.
Figure \ref{fig:deltat}a shows the distribution of the time differences for the whole period of the reference pulses, 10,136~ns, and Figure \ref{fig:deltat}b shows it for a subset of nine oscillations. The oscillations are well described by a periodic function with period of 54.79~ns, which is exactly the period of the trap RF. 

There are 185 fluorescence rate oscillations during the 10,136 ns cycle of the reference pulse. To sum the statistics in all oscillations, the distribution from Figure \ref{fig:deltat}a was folded into a single oscillation accounting for the measured period of 54.79~ns. This provides a composite distribution of the ion fluorescence rate during a single micromotion cycle, shown in Figure \ref{fig:deltat}c. It represents the probability of the fluorescence during a single RF cycle, and can be correlated with the measured position of the ions to determine the ion micromotion since the ion position time dependence is already measured.

Time-stamping of photons and external pulses is performed independently from each other during data collection, and the analysis described above is done offline during post-processing. The same time information for the ions would be accessible from the individual pixel time-stamps so a similar analysis could have been done without any need of external pulses. However, it is an efficient technique to accumulate the time difference statistics without prior knowledge of the oscillation frequency.  It allows to determine the oscillation period promptly, in real time, and can be used for fast online monitoring of the micromotion.

\section{Results}
\label{sec:results}

\begin{figure}[ht]
    \centering
    \includegraphics[width=.45\textwidth]{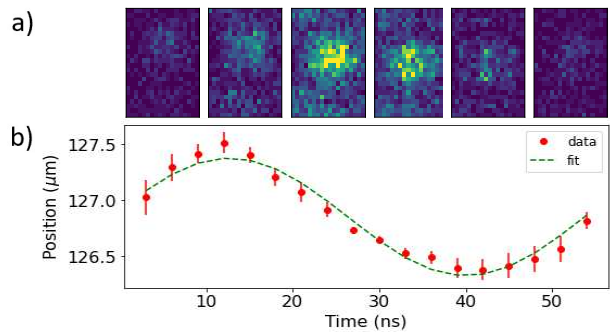}
    \centering
    \caption{Direct observation of the ion micromotion. a) Snapshots of the sixth ion during micromotion oscillation in six time bins separated by 9~ns each. The shown statistics corresponds to 3 ns time integration around the midpoint of the bin. b) Position of the sixth ion over time. The data is shown in red dots; the error bars come from fitting the ion image to a Gaussian function. The dashed green line is a sinusoidal fit to the data with a period $T = 54.79$~ns, amplitude $A =0.37 \pm 0.03$ $\mu$m and time offset $t_0 = 1.0 \pm 0.4$~ns.}
    \label{fig:positions}
\end{figure}

The presented optical setup allows imaging of ion motion in the plane perpendicular to the camera direction of view, spanned by the axis of the trap and the radial direction perpendicular to 
the optical axis of the camera. No significant micromotion was observed along the trap axis, since in this trap the RF field is mostly radial near the trap center. All further presented results correspond to ion motion along the radial direction. The measurement of the ion positions was performed by time-slicing the full oscillation period into multiple bins (``snapshots") and determining the ion coordinates in each time bin. As an example, Figure \ref{fig:positions}a shows such snapshots of the sixth ion from Figure \ref{fig:setup}b during the full period of the micromotion oscillation. Here six images are separated by 9~ns, and the image statistics corresponds to a 3~ns time integration at the time stamp of the bin. Both the displacement of the ion and the modulation of its fluorescence due to the Doppler shift are clearly visible. In the Supplemental Material at [URL will be inserted by publisher] the ``snapshot'' of 8 ions at 30 ns from of the 54.79 ns micromotion period as well as the animation of both the ions' displacements and fluorescence modulation due to the Doppler shift are provided. Figure \ref{fig:positions}b shows the measured y-position of the sixth ion as a function of the time in 3~ns bins. The data is well described by a sinusoidal function 
\begin{equation}
y = A \sin [2\pi (t + t_0) / T]
\end{equation}
with a period $T = 54.79$~ns, amplitude $A =0.37 \pm 0.03$ $\mu$m and time offset $t_0 = 1.0 \pm 0.4$~ns. The time offset is proportional to the phase of the micromotion. The period was determined as described in Section \ref{sec:analysis} while the amplitude and phase are fits to the data.

\begin{figure}[ht]
    \centering
    \includegraphics[width=8.5cm]{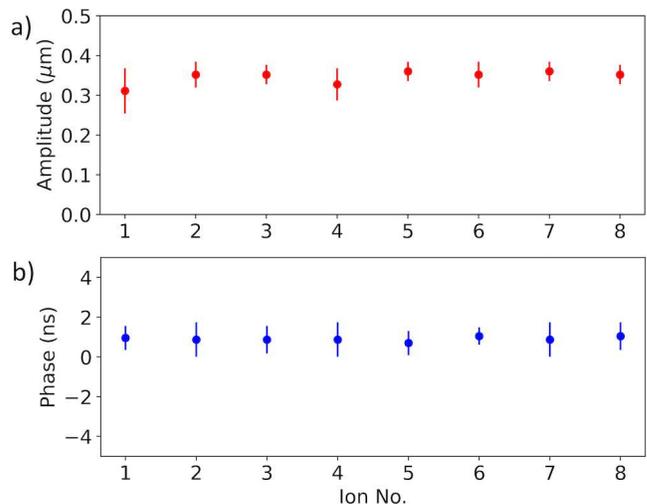}
    \centering
    \caption{Amplitudes (a) and time offsets (b) of the individual ions. Error bars are determined from the fit parameters in Equation~1.}
    \label{fig:Micromotion Data}
\end{figure}

The amplitudes and time offsets for all eight ions in the crystal are determined by performing the analysis individually for each ion. The results are shown in Figure \ref{fig:Micromotion Data}. All ions have similar values for amplitude and time offset, which agree well within the measurement uncertainties, implying their near identical micromotion. This indicates that the ion chain was shifted from the RF null of the trap without a tilt, as expected in our setup. An animation of the micromotion dynamics of the entire 8-ion chain is provided in the Supplemental Material by using ion chain snapshots.

\begin{figure}[ht]
    \centering
    \includegraphics[width=8.5cm]{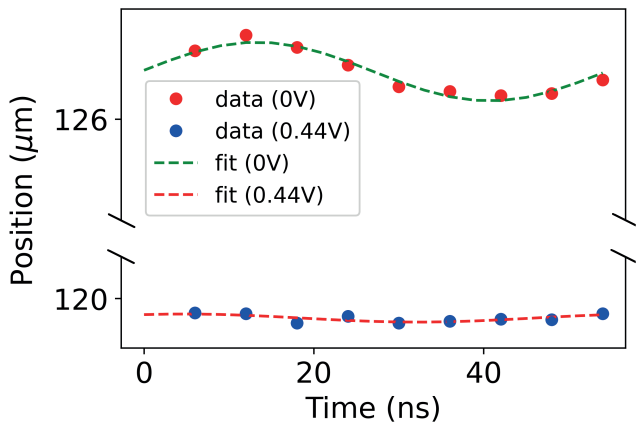}
    \centering
    \caption{\label{fig:compensated} The micromotion of the sixth ion without the DC compensation voltage (red dots with green dashed line fit) and with 0.44V, which corresponds to DC electric field of 550 V/m, applied (blue dots with red dashed line fit). The error bars are smaller than the dot size.
}

\end{figure}

By applying compensation DC voltage to the trap rods (Figure~\ref{fig:setup}a), the ion position along one of the radial axes can be shifted either toward or away from the trap RF null and, thus, can increase or decrease the micromotion amplitude in this direction. Figure \ref{fig:compensated} shows the micromotion of the sixth ion for two cases: without any DC compensation voltage (0V) and with 0.44V, which corresponds to 550 V/m, applied to one of the trap rods. In the latter case there is a considerable reduction of the micromotion amplitude from $0.37 \pm 0.03~\mu$m to $0.07 \pm 0.03 ~\mu$m. There is also a 7$~\mu$m displacement of the ion position caused by the applied voltage due to the changed electrostatic field configuration.

In addition to using the position information for micromotion measurement, time histograms like in Figure \ref{fig:deltat}c provide velocity information for each ion through the Doppler-shifted scattering rate. As ions undergo micromotion, the component of their velocity along the direction of the incident laser beam causes the frequency of the laser to be Doppler shifted in the reference frame of the ions proportional to the instantaneous velocity. The red-detuned Doppler cooling laser becomes less (more) detuned from the atomic resonance as the ion moves toward (away from) the laser, resulting in the increase (decrease) of the photon scattering rate~\cite{Berkeland98}. This modulation provides an independent way to estimate the component of micromotion along the direction of the laser beam, which can be used to estimate ion micromotion. This well-known technique has previously been implemented in many experiments by time-correlating single photons detected by a photomultiplier tube (PMT) with the RF phase using a TDC~\cite{Berkeland98,Pyka2014}. However, the PMT lacks spatial resolution so the old method only works for single ions. With the spatial resolution of the Tpx3Cam, precision determination of every ion's micromotion should be possible but here we limit our discussion to the novel direct micromotion detection technique.

Both techniques rely on the excellent time resolution of the camera, however  the direct measurement technique also relies on the precise reconstruction of ion positions, so it requires good coordinate resolution of the camera and a diffraction limited optical system with high light collection efficiency and low aberrations.

To investigate the limits of the presented technique, we estimate the minimum ion displacement 
that is possible to measure with this method. This provides the minimum detectable micromotion amplitude, which, in turn, can be used to calculate the minimum detectable stray DC electric field for any trap geometry, since it depends only on the spatial resolution of the camera. In this work, the upper limit of the spatial resolution can be estimated using the value of the standard deviation of the measured micromotion amplitude, which is about 50 nm. We believe that this limit could be improved by increasing the photon collection time, as well as by improving the quality of the ion image with a higher N.A. imaging optics.

\section{Conclusions}
\label{sec:conclusions}

This paper demonstrates the direct simultaneous measurement of the micromotion of eight ions in a linear RF trap using a single photon sensitive camera with nanosecond timing resolution that provides both time and position measurements for each detected photon. Amplitudes and time offsets of each ion's micromotion were determined and found to be consistent with a translation of the linear ion crystal away from the RF nodal line. Excess micromotion compensation was demonstrated to below $0.1 \mu$m. The new technique does not require prior knowledge of the cooling laser parameters or addressing narrow atomic transitions. No special modifications of the trap, lasers or imaging apparatus is necessary, since the same Tpx3Cam camera used in this work to observe the micromotion is also used as the regular imaging tool for setting up the trap, and can also be used as a fast and sensitive detector for high-fidelity trapped ion qubit state readout \cite{Zhukas2020}. In conjunction with a high numerical aperture objective lens, this method may be used to detect the micromotion component along the direction parallel to the imaging axis by analyzing the ion image defocusing~\cite{Gloger2015}, thus providing a full 3-dimensional picture of the micromotion. This technique is also applicable to more complex ion configurations such as two-dimensional ion crystals.

\begin{acknowledgments}
We thank Michael Keach and Kurt A. Delegard for assistance with the datasets, analysis and optical setup. This work was supported by the U.S. Department of Energy QuantISED award, U.S. National Science Foundation award PHY-2011503 and by the grant LM2018109 of Ministry of Education, Youth and Sports as well as by Centre of Advanced Applied Sciences CZ.02.1.01/0.0/0.0/16-019/0000778, co-financed by the European Union. M.M. acknowledges support under the Science Undergraduate Laboratory Internships Program (SULI) by the U.S. Department of Energy.
\end{acknowledgments}

\providecommand{\noopsort}[1]{}\providecommand{\singleletter}[1]{#1}%

\end{document}